\documentclass[12pt,a4paper]{article}

\usepackage{ifthen} 
\newboolean{pdflatex}
\setboolean{pdflatex}{true} 

\newboolean{articletitles}
\setboolean{articletitles}{true} 

\newboolean{uprightparticles}
\setboolean{uprightparticles}{false} 

\newboolean{inbibliography}
\setboolean{inbibliography}{false} 

\usepackage{microtype}
\usepackage{lineno}  
\usepackage{xspace} 
\usepackage{caption} 

\usepackage{graphicx}  
\usepackage{color}
\usepackage{colortbl}
\graphicspath{{./figs/}} 

\usepackage{amsmath} 
\usepackage{amssymb}
\usepackage{amsfonts}
\usepackage{upgreek} 

\newcommand*\patchAmsMathEnvironmentForLineno[1]{%
\expandafter\let\csname old#1\expandafter\endcsname\csname #1\endcsname
\expandafter\let\csname oldend#1\expandafter\endcsname\csname
end#1\endcsname
 \renewenvironment{#1}%
   {\linenomath\csname old#1\endcsname}%
   {\csname oldend#1\endcsname\endlinenomath}%
}
\newcommand*\patchBothAmsMathEnvironmentsForLineno[1]{%
  \patchAmsMathEnvironmentForLineno{#1}%
  \patchAmsMathEnvironmentForLineno{#1*}%
}
\AtBeginDocument{%
\patchBothAmsMathEnvironmentsForLineno{equation}%
\patchBothAmsMathEnvironmentsForLineno{align}%
\patchBothAmsMathEnvironmentsForLineno{flalign}%
\patchBothAmsMathEnvironmentsForLineno{alignat}%
\patchBothAmsMathEnvironmentsForLineno{gather}%
\patchBothAmsMathEnvironmentsForLineno{multline}%
\patchBothAmsMathEnvironmentsForLineno{eqnarray}%
}

\usepackage{hyperref}    
\usepackage[all]{hypcap} 

\def\MagUp {\mbox{\em Mag\kern -0.05em Up}\xspace}

\ifthenelse{\boolean{uprightparticles}}%
{

 \def\Pmu         {\ensuremath{\upmu}\xspace}

 \def\Ppi         {\ensuremath{\uppi}\xspace}

 \def\Ppsi        {\ensuremath{\uppsi}\xspace}

 \def\PDelta      {\ensuremath{\Delta}\xspace}                 
 \def\PXi      {\ensuremath{\Xi}\xspace}                 
 \def\PLambda      {\ensuremath{\Lambda}\xspace}                 
 \def\PSigma      {\ensuremath{\Sigma}\xspace}                 
 \def\POmega      {\ensuremath{\Omega}\xspace}                 
 \def\PUpsilon      {\ensuremath{\Upsilon}\xspace}                 
 
 \def\PB      {\ensuremath{\mathrm{B}}\xspace}                 
                  
 \def\PD      {\ensuremath{\mathrm{D}}\xspace}

 \def\PJ      {\ensuremath{\mathrm{J}}\xspace}                 
 \def\PK      {\ensuremath{\mathrm{K}}\xspace}

 \def\Pb      {\ensuremath{\mathrm{b}}\xspace}                 
 \def\Pc      {\ensuremath{\mathrm{c}}\xspace}

 \def\Pi      {\ensuremath{\mathrm{i}}\xspace}

 \def\Ps      {\ensuremath{\mathrm{s}}\xspace}

}
{

 \def\Pmu         {\ensuremath{\mu}\xspace}

 \def\Ppi         {\ensuremath{\pi}\xspace}

 \def\Ppsi        {\ensuremath{\psi}\xspace}                 
                  
 \mathchardef\PDelta="7101
 \mathchardef\PXi="7104
 \mathchardef\PLambda="7103
 \mathchardef\PSigma="7106
 \mathchardef\POmega="710A
 \mathchardef\PUpsilon="7107
                  
 \def\PB      {\ensuremath{B}\xspace}                 
                  
 \def\PD      {\ensuremath{D}\xspace}

 \def\PJ      {\ensuremath{J}\xspace}                 
 \def\PK      {\ensuremath{K}\xspace}

 \def\Pb      {\ensuremath{b}\xspace}                 
 \def\Pc      {\ensuremath{c}\xspace}

 \def\Pi      {\ensuremath{i}\xspace}

 \def\Ps      {\ensuremath{s}\xspace}

}

\makeatletter
\ifcase \@ptsize \relax
  \newcommand{\miniscule}{\@setfontsize\miniscule{4}{5}}
\or
  \newcommand{\miniscule}{\@setfontsize\miniscule{5}{6}}
\or
  \newcommand{\miniscule}{\@setfontsize\miniscule{5}{6}}
\fi
\makeatother

\DeclareRobustCommand{\optbar}[1]{\shortstack{{\miniscule (\rule[.5ex]{1.25em}{.18mm})}
  \\ [-.7ex] $#1$}}

\def\mumu       {{\ensuremath{\Pmu^+\Pmu^-}}\xspace}

\def\squark    {{\ensuremath{\Ps}}\xspace}

\def\cquark    {{\ensuremath{\Pc}}\xspace}

\def\bquark    {{\ensuremath{\Pb}}\xspace}

\def\pion   {{\ensuremath{\Ppi}}\xspace}

\def\pip    {{\ensuremath{\pion^+}}\xspace}
\def\pim    {{\ensuremath{\pion^-}}\xspace}

\def\kaon    {{\ensuremath{\PK}}\xspace}
  \def\Kbar    {{\kern 0.2em\overline{\kern -0.2em \PK}{}}\xspace}

\def\KorKbar    {\kern 0.18em\optbar{\kern -0.18em K}{}\xspace}

\def\Kp      {{\ensuremath{\kaon^+}}\xspace}
\def\Km      {{\ensuremath{\kaon^-}}\xspace}

  \def\Dbar    {{\kern 0.2em\overline{\kern -0.2em \PD}{}}\xspace}
\def\D       {{\ensuremath{\PD}}\xspace}

\def\DorDbar    {\kern 0.18em\optbar{\kern -0.18em D}{}\xspace}
\def\DtwoorDtwobar {\kern -0.25em\optbar{\kern 0.25em D_2^*}{}\xspace}
\def\Dz      {{\ensuremath{\D^0}}\xspace}

\def\Dp      {{\ensuremath{\D^+}}\xspace}

\def\B       {{\ensuremath{\PB}}\xspace}
\def\Bbar    {{\ensuremath{\kern 0.18em\overline{\kern -0.18em \PB}{}}}\xspace}

\def\BorBbar    {\kern 0.18em\optbar{\kern -0.18em B}{}\xspace}

\def\BzorBzbar  {\kern 0.18em\optbar{\kern -0.18em B}{}^0\xspace}

\def\Bub     {{\ensuremath{\B^-}}\xspace}

\def\Bm      {{\ensuremath{\Bub}}\xspace}

\def\Bsb     {{\ensuremath{\Bbar{}^0_\squark}}\xspace}

\def\Bcm     {{\ensuremath{\B_\cquark^-}}\xspace}

\def\jpsi     {{\ensuremath{{\PJ\mskip -3mu/\mskip -2mu\Ppsi\mskip 2mu}}}\xspace}

  \def\Y#1S{\ensuremath{\PUpsilon{(#1S)}}\xspace}

\def\Xires       {{\ensuremath{\PXi}}\xspace}

\def\Lz          {{\ensuremath{\PLambda}}\xspace}
\def\Lbar        {{\ensuremath{\kern 0.1em\overline{\kern -0.1em\PLambda}}}\xspace}
\def\LorLbar    {\kern 0.18em\optbar{\kern -0.18em \PLambda}{}\xspace}

\def\Lb      {{\ensuremath{\Lz^0_\bquark}}\xspace}

\def\Lc      {{\ensuremath{\Lz^+_\cquark}}\xspace}

\def\Xibm    {{\ensuremath{\Xires^-_\bquark}}\xspace}

\def\Xic     {{\ensuremath{\Xires_\cquark}}\xspace}

\def\Xicp    {{\ensuremath{\Xires^+_\cquark}}\xspace}

\def\to                 {\ensuremath{\rightarrow}\xspace}

\def\AT#1     {\ensuremath{A_{\mathrm{T}}^{#1}}\xspace}           

\def\C#1      {\ensuremath{\mathcal{C}_{#1}}\xspace}                       
\def\Cp#1     {\ensuremath{\mathcal{C}_{#1}^{'}}\xspace}                    
\def\Ceff#1   {\ensuremath{\mathcal{C}_{#1}^{\mathrm{(eff)}}}\xspace}        
\def\Cpeff#1  {\ensuremath{\mathcal{C}_{#1}^{'\mathrm{(eff)}}}\xspace}       
\def\Ope#1    {\ensuremath{\mathcal{O}_{#1}}\xspace}                       
\def\Opep#1   {\ensuremath{\mathcal{O}_{#1}^{'}}\xspace}                    

\newcommand{\tev}{\ifthenelse{\boolean{inbibliography}}{\ensuremath{~T\kern -0.05em eV}\xspace}{\ensuremath{\mathrm{\,Te\kern -0.1em V}}}\xspace}
\newcommand{\gev}{\ensuremath{\mathrm{\,Ge\kern -0.1em V}}\xspace}
\newcommand{\mev}{\ensuremath{\mathrm{\,Me\kern -0.1em V}}\xspace}
\newcommand{\kev}{\ensuremath{\mathrm{\,ke\kern -0.1em V}}\xspace}
\newcommand{\ev}{\ensuremath{\mathrm{\,e\kern -0.1em V}}\xspace}
\newcommand{\gevc}{\ensuremath{{\mathrm{\,Ge\kern -0.1em V\!/}c}}\xspace}
\newcommand{\mevc}{\ensuremath{{\mathrm{\,Me\kern -0.1em V\!/}c}}\xspace}
\newcommand{\gevcc}{\ensuremath{{\mathrm{\,Ge\kern -0.1em V\!/}c^2}}\xspace}
\newcommand{\gevgevcccc}{\ensuremath{{\mathrm{\,Ge\kern -0.1em V^2\!/}c^4}}\xspace}
\newcommand{\mevcc}{\ensuremath{{\mathrm{\,Me\kern -0.1em V\!/}c^2}}\xspace}

\def\mum  {\ensuremath{{\,\upmu\rm m}}\xspace}

\def\nb {\ensuremath{\rm \,nb}\xspace}
\def\invnb {\ensuremath{\mbox{\,nb}^{-1}}\xspace}

\def\invfb   {\ensuremath{\mbox{\,fb}^{-1}}\xspace}

\def\ps   {\ensuremath{{\rm \,ps}}\xspace}

\newcommand{\chisq}{\ensuremath{\chi^2}\xspace}

\def\gsim{{~\raise.15em\hbox{$>$}\kern-.85em
          \lower.35em\hbox{$\sim$}~}\xspace}
\def\lsim{{~\raise.15em\hbox{$<$}\kern-.85em
          \lower.35em\hbox{$\sim$}~}\xspace}

\def\pt         {\mbox{$p_{\rm T}$}\xspace}

\def\tell1  {TELL1\xspace}
\def\ukl1   {UKL1\xspace}

\newcommand{\ie}{\mbox{\itshape i.e.}\xspace}

\usepackage{cite} 
\usepackage{mciteplus}

\usepackage{rotating} 
\usepackage{arydshln} 
\usepackage{relsize} 
\usepackage[hang,flushmargin]{footmisc}

\begin{document}

\renewcommand{\thefootnote}{\fnsymbol{footnote}}
\setcounter{footnote}{1}

\begin{titlepage}
\pagenumbering{roman}

{\bf\boldmath\huge
\begin{center}
  Displaced \Bcm\ mesons as an inclusive signature of weakly decaying double beauty hadrons
\end{center}
}

\vspace*{1.5cm}

\begin{center}
  T.~Gershon$^1$, A.~Poluektov$^{1,2}$
\bigskip\\
{\it\footnotesize 
$ ^1$ Department of Physics, University of Warwick, Coventry, United Kingdom\\
$ ^2$ Aix Marseille Univ, CNRS/IN2P3, CPPM, Marseille, France\\
}
\end{center}

\vspace{\fill}

\begin{abstract}
  \noindent
  The recent discovery, by the LHCb collaboration, of the $\PXi_{cc}^{++}$ doubly charmed baryon, has renewed interest in the spectroscopy of doubly heavy hadrons.
  Experimentally, however, searches for such states appear highly challenging.
  The reconstructed final states tend to involve multiple heavy flavoured (beauty or charm) hadrons, so the yield for any exclusive decay mode will be suppressed to unobservably low levels by the product of several branching fractions, each of which is typically $10^{-3}$--$10^{-2}$.
  Noting that decays of double beauty hadrons are the only possible source of \Bcm\ mesons that are displaced from the primary vertices of proton-proton collisions at the LHC, a more promising inclusive search strategy is proposed.
\end{abstract}

\vspace{\fill}

\end{titlepage}

\newpage
\setcounter{page}{2}
\mbox{~}

\cleardoublepage

\renewcommand{\thefootnote}{\arabic{footnote}}
\setcounter{footnote}{0}

\pagestyle{plain} 
\setcounter{page}{1}
\pagenumbering{arabic}

The possible existence of doubly heavy baryons, \ie\ bound states that contain two or more beauty or charm quarks, has been predicted since the discoveries of those quarks, not long after the postulation of the quark model.
However, such states have relatively low production cross-sections, and require sophisticated detectors to be able to distinguish their decays from backgrounds.
The first definitive observation of a doubly heavy baryon was reported by the LHCb collaboration in 2017, with a clear signal of $\PXi_{cc}^{++} \to \Lc\Km\pip\pip$ decays obtained from analysis of a data sample of high-energy proton-proton collisions~\cite{LHCb-PAPER-2017-018}.
Subsequently, the $\Xicp\pip$ decay mode has also been observed~\cite{LHCb-PAPER-2018-026}, and the lifetime of the $\PXi_{cc}^{++}$ state has been found to be consistent with expectation for decays mediated by the weak interaction~\cite{LHCb-PAPER-2018-019}.

This discovery has prompted various theoretical predictions for the production and decay properties of other doubly heavy states.
Interestingly, these have not been confined to baryonic states.
Various recently discovered ``exotic'' hadrons, which do not fit into the conventional scheme of $q\bar{q}$ mesons and $qqq$ baryons, have led to several different interpretations including models based on tightly bound tetraquarks (see, for example, Refs.~\cite{Olsen:2017bmm,Ali:2017jda,Lebed:2016hpi} for reviews).
Several authors have predicted the existence of a $bb\bar{u}\bar{d}$ tetraquark state with mass below the threshold for decay to a pair of beauty mesons~\cite{Ader:1981db,Manohar:1992nd,Bicudo:2015vta,Francis:2016hui,Karliner:2017qjm,Eichten:2017ffp,Czarnecki:2017vco}.\footnote{
  Some of these predictions predate the $\PXi_{cc}^{++}$ discovery, in two cases quite considerably so.
  The experimental data have, however, put the theory on much firmer ground.
}
It is possible that the $bb\bar{u}\bar{s}$ and $bb\bar{d}\bar{s}$ tetraquarks are also stable against strong and electromagnetic decays.
Since these tetraquarks can only decay weakly, they are expected to have comparable lifetimes to the ground state double beauty baryons, \ie\ $\sim 0.4$--$0.8 \ps$~\cite{Kiselev:2001fw,Karliner:2014gca,Berezhnoy:2018bde}.

Double beauty hadrons $bbx$, where $x$ is either a quark or an anti-diquark, can be produced in high energy collisions such as those at the LHC where, assuming non-negligible lifetimes, they will travel a measurable distance before decaying.
Typical estimates for the production cross-sections are ${\cal O}(1 \nb)$~\cite{Ali:2018xfq}. 
Since the accumulated data sample at LHCb (ATLAS and CMS) already exceeds $8\,(150)\invfb$, with increases of over an order of magnitude anticipated in the HL-LHC era, it might be assumed that prospects for discovery are good.
Unfortunately, branching fractions for heavy flavour decays to even the most abundant final states are typically $10^{-2}$--$10^{-3}$.
Exclusive reconstruction of double beauty hadrons will inevitably involve a chain of such decays.
It is also relevant that the final states will involve many particles, leading to lower reconstruction efficiencies. 
Consequently, the yields that can be obtained are expected to be suppressed to unobservably low levels.

\begin{figure}[!tb]
  \centering
  \includegraphics[width=0.58\textwidth]{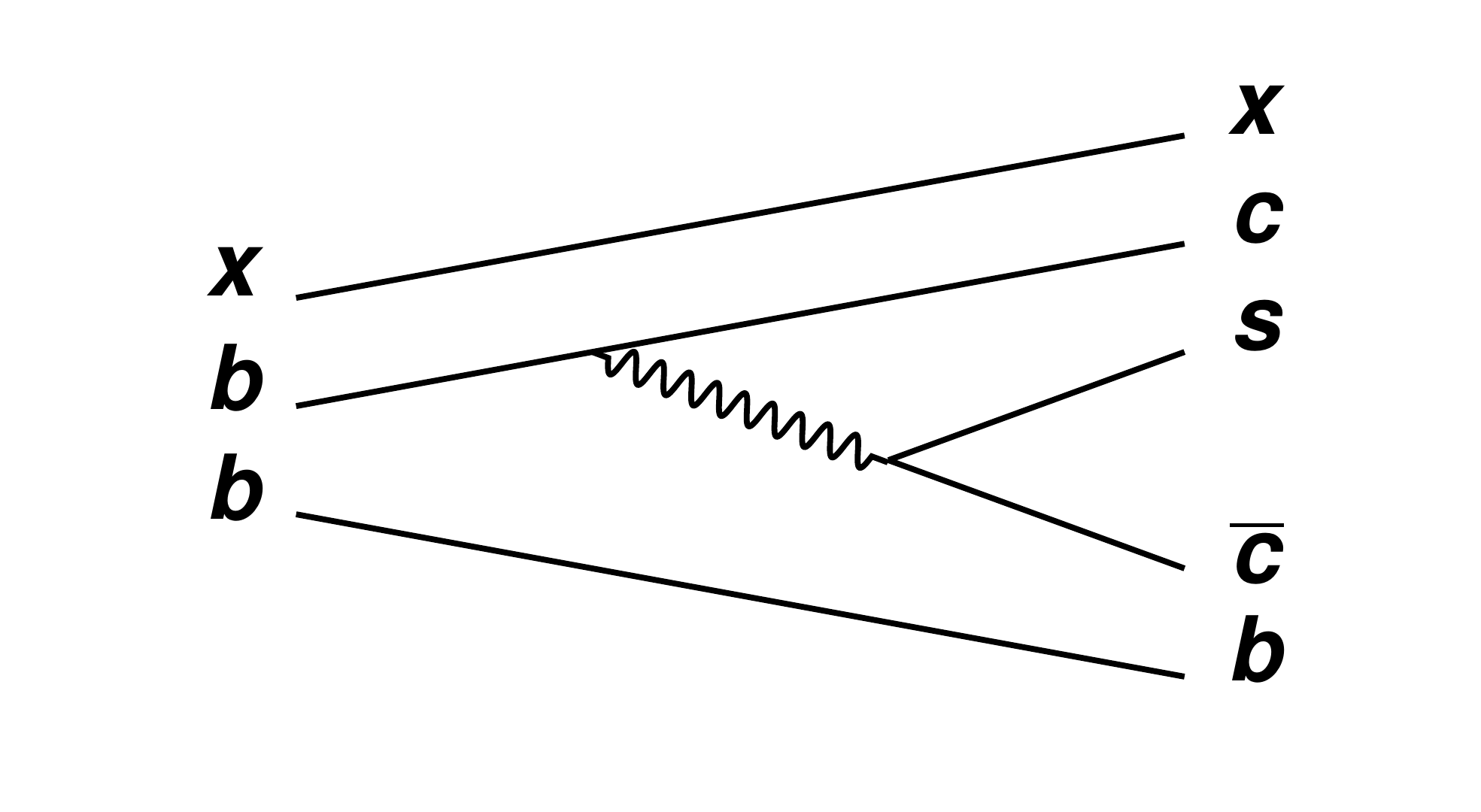}
  \caption{
    Diagram for production of a $\Bcm$ meson from a double beauty hadron decay.
  }
  \label{fig:diagram}
\end{figure}

This motivates the development of inclusive search strategies. 
An important realisation, in this respect, is that weak decays of double beauty hadrons are the only possible source of \Bcm\ mesons that are displaced from, \ie\ do not originate directly from, the primary vertices of the LHC collisions.
This is a simple consequence of the facts that: (i) all fundamental Standard Model particles heavier than the $b$ quark decay instantaneously, and do not produce a displaced vertex; (ii) any hadron containing both $b$ and $\bar{b}$ constituent quarks will decay via strong and electromagnetic interactions and hence not have a measurable lifetime. 
Displaced \Bcm\ mesons can therefore only be produced\footnote{
    Another conceivable source of displaced \Bcm\ mesons would be from weak decays of triply heavy $bq\bar{c}\bar{c}$ tetraquark states, where $q \in u,d$.
    However, even if such a state is stable against strong decays, its production rate will be tiny, its branching fraction for decays to final states involving \Bcm\ mesons will be small, and its lifetime will be very short.
    This potential contribution is therefore considered to have  negligible impact on the proposed measurement strategy.
} when one of the beauty quarks in a $bbx$ hadron decays via a $b \to \bar{c}$ transition and the produced $\bar{c}$ antiquark hadronises with the remaining $b$ quark as shown in Fig.~\ref{fig:diagram}.  
The $b \to c\bar{c}s$ transition will be dominant, since amplitudes for $b \to c\bar{c}d$, $b \to u \bar{c}s$ and $b \to u \bar{c}d$ processes involve smaller CKM matrix elements.
The inclusion of charge conjugate processes is implied throughout this paper.

The signature of displaced \Bcm\ mesons is quite attractive experimentally, especially for the LHCb experiment which can exploit its precision vertex detector to separate signal from (prompt) background~\cite{Alves:2008zz,LHCb-DP-2014-001}.
Moreover, when the $\Bcm \to \jpsi \pim, \jpsi \to \mumu$ decay chain is used there are only three final state particles to be reconstructed, and the $\jpsi \to \mumu$ signal helps to suppress backgrounds, so that high efficiency can be achieved.
The branching fraction for this \Bcm\ decay is also relatively large, typically expected to be a few percent~\cite{Qiao:2012hp}.
Its exact value is not known since the \Bcm\ production cross-section is unmeasured.
The ratio
$$
\frac{f_c}{f_u}\frac{{\cal B}\left(\Bcm \to \jpsi\pim\right)}{{\cal B}\left(\Bm \to \jpsi\Km\right)} = (0.683 \pm 0.018 \pm 0.009)\%
$$
has been measured by LHCb in the fiducial range $\pt(\B) < 20 \gev$ and $2.0 < y(\B) < 4.5$~\cite{LHCb-PAPER-2014-050} (units in which $c=1$ are used throughout this paper). 
Here, $f_c$ and $f_u$ are the fragmentation fractions for \Bcm\ and \Bm\ mesons, respectively, while \pt\ and $y$ denote the component of momentum transverse to the beam direction and the rapidity.
This result and the prediction for ${\cal B}\left(\Bcm \to \jpsi\pim\right)$ are consistent with the expectation that $\frac{f_c}{f_u} \sim {\cal O}\left(10^{-3}\right)$.
In spite of this low production rate, a yield of $3325 \pm 73$ $\Bcm \to \jpsi \pim$ decays has been obtained in $2 \invfb$ of proton-proton collision data at centre-of-mass energy $\sqrt{s} = 8 \tev$, recorded at LHCb~\cite{LHCb-PAPER-2017-042}.
This is significantly larger than the yields reconstructed in other decay modes, and the $\Bcm \to \jpsi \pim$ channel will therefore be the focus of this study.
The $\Bcm \to \jpsi \pim\pip\pim$ mode~\cite{LHCB-PAPER-2011-044}, with a yield about half as large as that for $\Bcm \to \jpsi \pim$, may also contribute useful sensitivity to an experimental analysis.

The expected yield of displaced \Bcm\ mesons is given by the product of the following factors, where order-of-magnitude estimated (indicated by "$\sim$") or known values are given in parentheses:
the $bbx$ hadron production cross-section ($\sim 1 \nb$), the branching fractions for its inclusive decay to final states containing \Bcm\ mesons ($\sim 10\%$) and for the subsequent $\Bcm \to \jpsi\pim$ ($\sim 2\%$) and $\jpsi \to \mumu$ (6\%) transitions, the detection efficiency ($\sim 10\%$) and the integrated luminosity.
Thus, around 10 displaced \Bcm\ mesons per $\invfb$ may be reconstructed.
To put it another way, ${\cal O}(1\%)$ of all \Bcm\ mesons detected in experiments at the Large Hadron Collider may originate from decays of double beauty hadrons.

The signature of displaced meson production has been used in various ways to determine heavy flavour hadron production cross-sections.
For example, displaced \jpsi\ mesons are commonly used as a signature of \bquark\ hadron decays -- all LHC experiments have exploited this approach to measure the $b\bar{b}$ production rate in different collision environments. 
In these analyses, the prompt and displaced \jpsi\ mesons are typically separated by fitting the distribution of a pseudo-proper decay time variable,
$$
t_z = \frac{(z_\jpsi - z_{\rm PV}) \times M_\jpsi}{p_{z\,\jpsi}} \, ,
$$
where $z_\jpsi$ and $z_{\rm PV}$ are the respective positions of the \jpsi\ decay vertex and the primary vertex, and $p_{z\,\jpsi}$ is the component of the \jpsi\ momentum, along the beam direction.
This approach cannot however be used to study displaced \Bcm\ mesons, as the non-negligible \Bcm\ lifetime of $\sim 0.5 \ps$~\cite{LHCb-PAPER-2013-063,LHCb-PAPER-2014-060} causes the production and decay vertices to be separated.

\begin{figure}[!tb]
  \centering
  \includegraphics[width=0.67\textwidth]{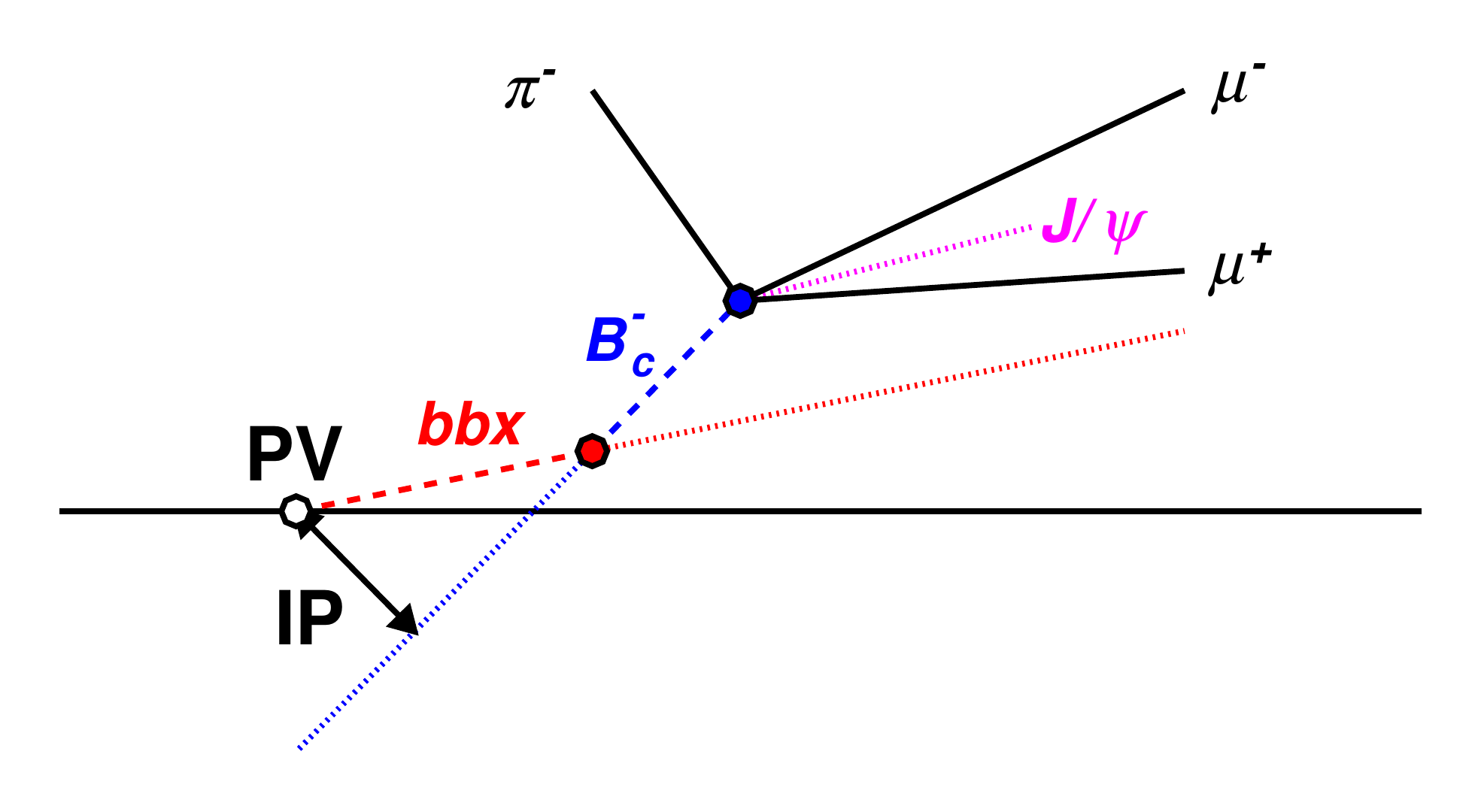}
  \caption{
    Illustration of the non-zero IP associated with $\Bcm$ production from weakly decaying $bbx$ hadrons.
    Additional particles produced in the $bbx$ decay are not shown.  
    Reconstruction of the $\Bcm$ decay vertex position and momentum vector from the $\jpsi\pim$ final state particles alone is sufficient to determine the IP.
  }
  \label{fig:cartoon}
\end{figure}

\begin{figure}[!tb]
  \centering
  \includegraphics[width=0.5\textwidth]{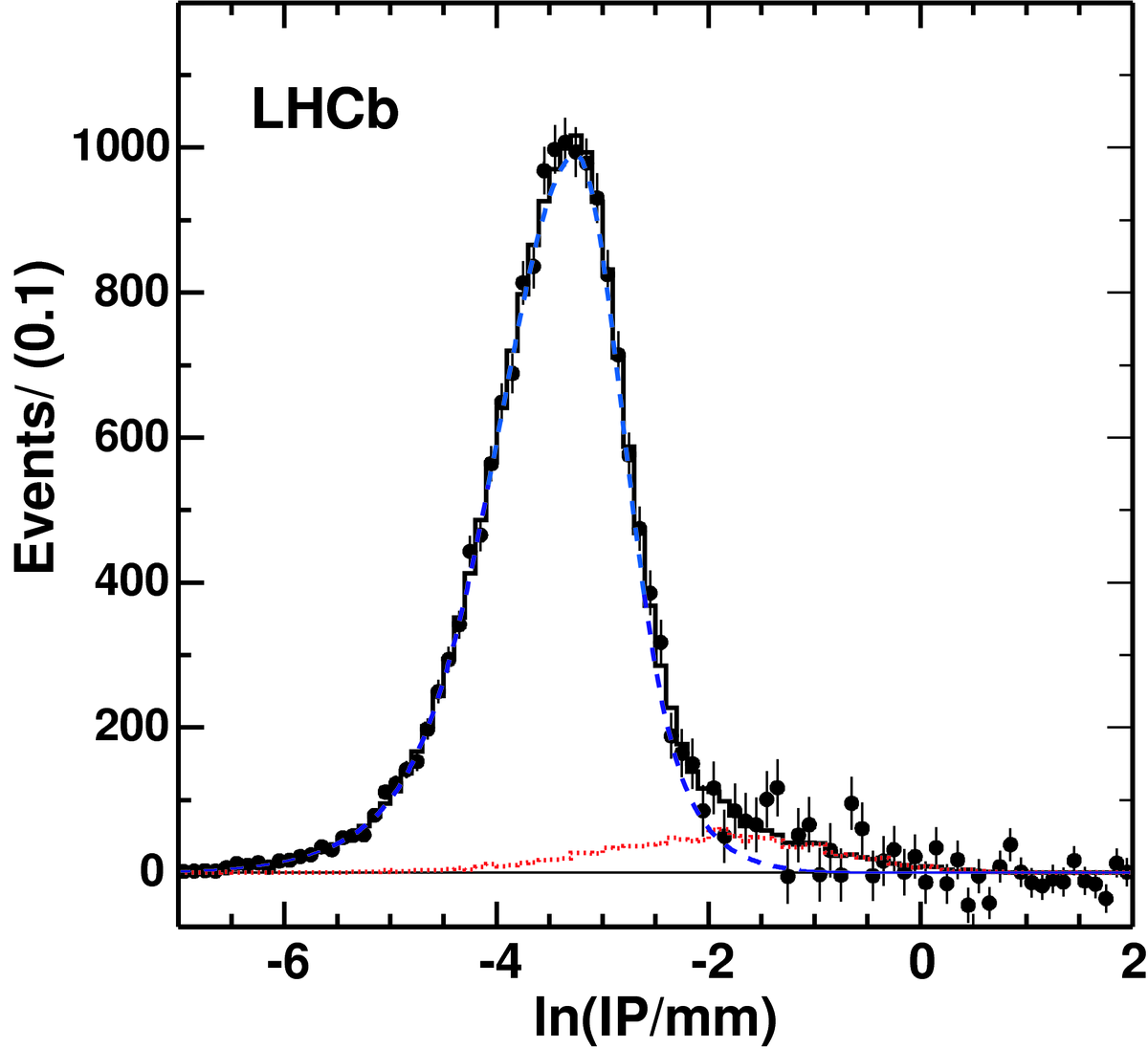}
  \caption{
    Separation of prompt and displaced $D^0$ mesons using the IP variable~\cite{LHCb-PAPER-2010-002}.
    The data correspond to $2.9 \invnb$ of $pp$ collisions, and contain a large prompt component (dashed line) and a smaller displaced component (dotted line).
    The sum of the two components is shown by a histogram.
  }
  \label{fig:D0IP}
\end{figure}

An alternative variable, the impact parameter (IP) can be used to identify displaced $\Bcm$ mesons, as illustrated in Fig.~\ref{fig:cartoon}.
Due to the non-negligible lifetime of the $bbx$ hadron, the momentum vector of the produced $\Bcm$ meson will not point back directly at the primary vertex, \ie\ the IP will be non-zero.
The use of the IP to separate displaced and prompt production has previously been used to identify displaced charm hadrons produced in \bquark\ hadron decays~\cite{LHCb-PAPER-2010-002,LHCb-PAPER-2011-018,LHCb-PAPER-2016-031}.\footnote{
  In Refs.~\cite{LHCb-PAPER-2010-002,LHCb-PAPER-2011-018,LHCb-PAPER-2016-031}, but not in Fig.~\ref{fig:D0IP}, the large branching fraction for $b \to c \mu\nu$ semileptonic decays is exploited and a muon is required to be associated to the charm hadron, which significantly suppresses prompt charm.
  This approach would not be possible in analysis of displaced \Bcm\ mesons, which are produced through $b \to \bar{c}$ transitions.
}
As shown in Fig.~\ref{fig:D0IP}, the displaced component is clearly visible at higher IP values, despite the relative yields of displaced and prompt $\Dz$ mesons being approximately $1:20$.
This signal-to-background ratio is somewhat more favourable than expected for displaced \Bcm\ mesons, however better alignment has led to improvement of the LHCb IP resolution since the analysis shown in Fig.~\ref{fig:D0IP} was performed, and further improvement is expected with the upgraded vertex locator to be installed before data taking in 2021~\cite{LHCb-DP-2014-001,LHCb-TDR-013}.

In addition to the lifetime and momentum of the weakly decaying double beauty hadron, the shape of the IP distribution for displaced \Bcm\ mesons depends on the momentum released in the decay (\ie\ the value $Q$, which is the difference between the  decaying hadron mass and the sum of the final state particle masses).
In the limit of small $Q$ value, the momentum vectors of the \Bcm\ meson and its parent particle will be collinear, and hence small IP values will be obtained.
For double beauty baryon decays producing a \Bcm\ meson through a $b \to c\bar{c}s$ transition, the final state must contain a charm-strange baryon ($m_{\Xic} \sim 2.47 \gev$), while for double beauty tetraquark decays a $DK$ combination ($m_D + m_K \sim 2.36 \gev$) must be present.  
Thus, a similar maximum $Q$ value is obtained in both cases: assuming a $bbx$ mass of $10.5 \gev$, $Q_{\rm max} \sim 1.8 \gev$.
This is sufficient to obtain a significantly non-zero IP value.

In order to understand how the unknown properties of the $bbx$ hadrons, in particular their masses and lifetimes, affect the IP distribution and hence the feasibility of the proposed analysis approach, a study based on a simple simulation of $bb\bar{u}\bar{d}$ tetraquarks has been performed. 
The kinematic properties, within the LHCb acceptance, are assumed to be the same as those for $B$ mesons, \ie\ approximately flat in pseudorapidity and having an exponential distribution in \pt\ with mean $5 \gev$~\cite{LHCb-PAPER-2017-037}.
Decay to the $\Bcm \Dp\Km$ final state, uniform over phase space, is assumed, with subsequent $\Bcm \to \jpsi\pim$, $\jpsi \to \mumu$ transitions.
Transverse momentum requirements, similar to those in other studies of $\Bcm$ mesons at LHCb~\cite{LHCb-PAPER-2017-042}, are imposed on the final-state particles. 

\begin{figure}[!tb]
  \centering
  \includegraphics[width=0.7\textwidth]{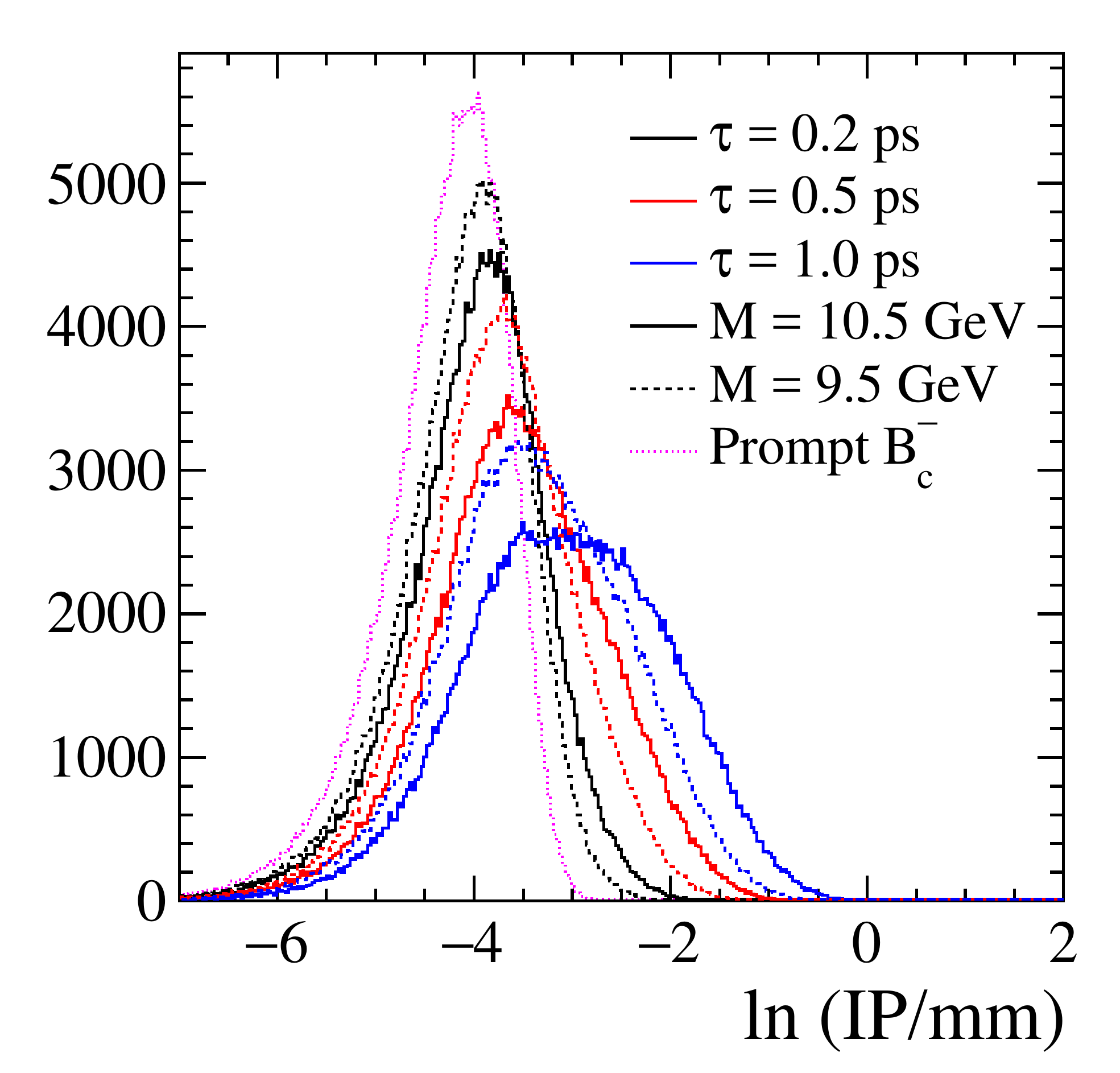} 
  \caption{
    IP distributions obtained from simplified simulations for different $bbx$ hadron masses (and thus $Q$ values) and lifetimes.
  }
  \label{fig:IPshapes}
\end{figure}

Within an experimental analysis, the $\Bcm$ IP would be determined from a full vertex fit of the three tracks, accounting for the covariance matrices of their position and momentum measurements, which is beyond the scope of the current study.
Instead, the IP resolution is assumed to be dominated by the coordinate resolution of the $\Bcm$ vertex, so that the IP resolution can be obtained as the uncertainty of the weighted average of the IP values for the three tracks. 
The per-track IP resolution is taken to be that obtained by LHCb with data collected in 2012, \ie\ $\sigma({\rm IP}_{x,y})~[\mum] = 11.6 + 23.4/(\pt~[\gev])$~\cite{LHCb-DP-2014-001}. 
The shapes obtained for a range of relevant parameters are shown in Fig.~\ref{fig:IPshapes}.
It can be seen that, for $bbx$ lifetimes above about $0.5 \ps$, the distribution for displaced $\Bcm$ hadrons extends to significantly higher IP values than that for prompt production.

The shape of the IP distribution for prompt \Bcm\ mesons is given purely by the resolution function, since the underlying true distribution is a delta function.
Given the expected signal-to-background ratio of around $1:100$, the sensitivity of the analysis will depend critically on the detector IP resolution -- it must be as precise as possible to separate prompt and displaced production, and must be well-known to avoid potentially large systematic uncertainties.
It should be stressed that the IP distribution for prompt $\Bcm$ mesons shown in Fig.~\ref{fig:IPshapes} is based on a simple study containing several approximations, and therefore is not expected to provide an accurate description of the real experimental situation, particularly in the tails.
It will be crucial in an experimental analysis that a suitable control mode can be used to study the resolution function.
Fortunately, the IP resolution for $\Bcm \to \jpsi\pim$ decays can be studied in data using the $\Bm \to \jpsi\Km$ decay as a control sample.
Contributions from non-prompt $\Bm$ mesons can be reduced by requiring production in narrow $\Bbar{}_s^{**0} \to \Bm \Kp$ decays~\cite{Aaltonen:2007ah,LHCb-PAPER-2012-030} where the associated $\Kp$ track is required to originate from the primary vertex.
Therefore, systematic uncertainties due to the tail of the resolution function should be controllable.

While the IP is a powerful variable that illustrates how prompt and displaced $\Bcm$ mesons can be separated, there is also additional information that could potentially be exploited to improve the sensitivity in an experimental analysis.
The better separation of prompt and displaced components at higher transverse momentum could be exploited by studying the two-dimensional distribution of IP and \pt.
The per-candidate estimated uncertainty on the IP measurement could alternatively be used in the analysis to enhance the sensitivity; at LHCb, this is usually done by constructing a $\chisq$ variable.  
Similarly, better separation is expected at larger values of the distance between the PV and the $\Bcm$ decay vertex.
Determining the optimal analysis approach would be an important aspect of experimental study.

It must be noted that whilst the signature of displaced \Bcm\ mesons would be sufficient to indicate double beauty hadron production, it would provide only limited information about the properties of those hadrons.
The yield is related to the production cross-section times the inclusive branching fraction for decay to final states containing $\Bcm$ mesons, summed over all weakly decaying $bbx$ hadrons.
Therefore, with improved predictions for these cross-sections and branching fractions, it could be possible to test whether only double beauty baryons are contributing, or if there is evidence for double beauty tetraquark production.
The shape of the distribution depends on the mass, lifetime and kinematic properties of the $bbx$ hadrons, averaged over all the contributing species.
Hence with predictions for two of these, the third could potentially be inferred.
Moreover, an observation of $bbx$ hadron production from an inclusive analysis would be likely to lead to insights as to how further information could be obtained, for example on which exclusive decays may be most accessible.

Finally, it is worthwhile to consider if any other inclusive signatures of displaced heavy flavour hadrons can be used similarly.
Secondary $B$ mesons can only be produced in decays of doubly heavy hadrons, with beauty charm hadron decays likely to be the dominant source.
Indeed, the $\Bcm \to \Bsb\pim$ decay has already been observed~\cite{LHCb-PAPER-2013-044}.
However, the $Q$ value for $B$ meson production in \Bcm\ or $\PXi_{bc}$ decays is relatively small, making it harder to separate signal from prompt production through the IP distribution.  
This effect also exists, but is less pronounced, for $b$ baryon production, so a search for displaced \Lb\ baryons as an inclusive signature of beauty charm hadrons could potentially be interesting.
(Secondary \Lb\ production through $\Xibm \to \Lb\pim$ is possible~\cite{LHCb-PAPER-2015-047} but has a very low $Q$ value and hence no significant contribution to the IP.)
Another exciting possibility for the long term is that displaced $\PXi_{cc}$ baryons could also be used to search for these states.
In these cases a reconstructed muon produced through semileptonic decay could potentially be used to tag the signal and reduce the prompt background significantly.

In conclusion, displaced \Bcm\ mesons provide a distinctive signature of the production of double beauty hadrons.
There appear to be reasonable prospects for the detection of such a signal with the large data samples accumulated by the LHCb detector and its planned upgrades, since excellent IP resolution is a feature of this experiment. 

\section*{Acknowledgements}

The authors wish to thank their colleagues on the LHCb experiment for the fruitful and enjoyable collaboration that inspired this study.
In particular, they would like to thank Vanya Belyaev, Steve Blusk, Alex Bondar, Mat Charles, Vladimir Gligorov, Jibo He and Mika Vesterinen for helpful comments.
This work is supported by the Science and Technology Facilities Council.

\addcontentsline{toc}{section}{References}
\setboolean{inbibliography}{true}
\bibliographystyle{LHCb}
\bibliography{references,main,LHCb-PAPER,LHCb-CONF,LHCb-DP,LHCb-TDR}

\ifx\mcitethebibliography\mciteundefinedmacro
\PackageError{LHCb.bst}{mciteplus.sty has not been loaded}
{This bibstyle requires the use of the mciteplus package.}\fi
\providecommand{\href}[2]{#2}
\begin{mcitethebibliography}{10}
\mciteSetBstSublistMode{n}
\mciteSetBstMaxWidthForm{subitem}{\alph{mcitesubitemcount})}
\mciteSetBstSublistLabelBeginEnd{\mcitemaxwidthsubitemform\space}
{\relax}{\relax}

\bibitem{LHCb-PAPER-2017-018}
LHCb collaboration, R.~Aaij {\em et~al.},
  \ifthenelse{\boolean{articletitles}}{\emph{{Observation of the doubly charmed
  baryon $\Xires_{cc}^{++}$}},
  }{}\href{https://doi.org/10.1103/PhysRevLett.119.112001}{Phys.\ Rev.\ Lett.\
  \textbf{119} (2017) 112001},
  \href{http://arxiv.org/abs/1707.01621}{{\normalfont\ttfamily
  arXiv:1707.01621}}\relax
\mciteBstWouldAddEndPuncttrue
\mciteSetBstMidEndSepPunct{\mcitedefaultmidpunct}
{\mcitedefaultendpunct}{\mcitedefaultseppunct}\relax
\EndOfBibitem
\bibitem{LHCb-PAPER-2018-026}
LHCb collaboration, R.~Aaij {\em et~al.},
  \ifthenelse{\boolean{articletitles}}{\emph{{First observation of the doubly
  charmed baryon decay \decay{\Xires_{cc}^{++}}{\Xires_c^+\pi^+} decay}},
  }{}\href{https://doi.org/10.1103/PhysRevLett.121.162002}{Phys.\ Rev.\ Lett.\
  \textbf{121} (2018) 162002},
  \href{http://arxiv.org/abs/1807.01919}{{\normalfont\ttfamily
  arXiv:1807.01919}}\relax
\mciteBstWouldAddEndPuncttrue
\mciteSetBstMidEndSepPunct{\mcitedefaultmidpunct}
{\mcitedefaultendpunct}{\mcitedefaultseppunct}\relax
\EndOfBibitem
\bibitem{LHCb-PAPER-2018-019}
LHCb collaboration, R.~Aaij {\em et~al.},
  \ifthenelse{\boolean{articletitles}}{\emph{{Measurement of the lifetime of
  the doubly charmed baryon $\Xires_{cc}^{++}$}},
  }{}\href{https://doi.org/10.1103/PhysRevLett.121.052002}{Phys.\ Rev.\ Lett.\
  \textbf{121} (2018) 052002},
  \href{http://arxiv.org/abs/1806.02744}{{\normalfont\ttfamily
  arXiv:1806.02744}}\relax
\mciteBstWouldAddEndPuncttrue
\mciteSetBstMidEndSepPunct{\mcitedefaultmidpunct}
{\mcitedefaultendpunct}{\mcitedefaultseppunct}\relax
\EndOfBibitem
\bibitem{Olsen:2017bmm}
S.~L. Olsen, T.~Skwarnicki, and D.~Zieminska,
  \ifthenelse{\boolean{articletitles}}{\emph{{Nonstandard heavy mesons and
  baryons: Experimental evidence}},
  }{}\href{https://doi.org/10.1103/RevModPhys.90.015003}{Rev.\ Mod.\ Phys.\
  \textbf{90} (2018) 015003},
  \href{http://arxiv.org/abs/1708.04012}{{\normalfont\ttfamily
  arXiv:1708.04012}}\relax
\mciteBstWouldAddEndPuncttrue
\mciteSetBstMidEndSepPunct{\mcitedefaultmidpunct}
{\mcitedefaultendpunct}{\mcitedefaultseppunct}\relax
\EndOfBibitem
\bibitem{Ali:2017jda}
A.~Ali, J.~S. Lange, and S.~Stone,
  \ifthenelse{\boolean{articletitles}}{\emph{{Exotics: Heavy pentaquarks and
  tetraquarks}}, }{}\href{https://doi.org/10.1016/j.ppnp.2017.08.003}{Prog.\
  Part.\ Nucl.\ Phys.\  \textbf{97} (2017) 123},
  \href{http://arxiv.org/abs/1706.00610}{{\normalfont\ttfamily
  arXiv:1706.00610}}\relax
\mciteBstWouldAddEndPuncttrue
\mciteSetBstMidEndSepPunct{\mcitedefaultmidpunct}
{\mcitedefaultendpunct}{\mcitedefaultseppunct}\relax
\EndOfBibitem
\bibitem{Lebed:2016hpi}
R.~F. Lebed, R.~E. Mitchell, and E.~S. Swanson,
  \ifthenelse{\boolean{articletitles}}{\emph{{Heavy-quark QCD exotica}},
  }{}\href{https://doi.org/10.1016/j.ppnp.2016.11.003}{Prog.\ Part.\ Nucl.\
  Phys.\  \textbf{93} (2017) 143},
  \href{http://arxiv.org/abs/1610.04528}{{\normalfont\ttfamily
  arXiv:1610.04528}}\relax
\mciteBstWouldAddEndPuncttrue
\mciteSetBstMidEndSepPunct{\mcitedefaultmidpunct}
{\mcitedefaultendpunct}{\mcitedefaultseppunct}\relax
\EndOfBibitem
\bibitem{Ader:1981db}
J.~P. Ader, J.~M. Richard, and P.~Taxil,
  \ifthenelse{\boolean{articletitles}}{\emph{{Do narrow heavy multi-quark
  states exist?}}, }{}\href{https://doi.org/10.1103/PhysRevD.25.2370}{Phys.\
  Rev.\  \textbf{D25} (1982) 2370}\relax
\mciteBstWouldAddEndPuncttrue
\mciteSetBstMidEndSepPunct{\mcitedefaultmidpunct}
{\mcitedefaultendpunct}{\mcitedefaultseppunct}\relax
\EndOfBibitem
\bibitem{Manohar:1992nd}
A.~V. Manohar and M.~B. Wise,
  \ifthenelse{\boolean{articletitles}}{\emph{{Exotic $QQ \bar{q}\bar{q}$ states
  in QCD}}, }{}\href{https://doi.org/10.1016/0550-3213(93)90614-U}{Nucl.\
  Phys.\  \textbf{B399} (1993) 17},
  \href{http://arxiv.org/abs/hep-ph/9212236}{{\normalfont\ttfamily
  arXiv:hep-ph/9212236}}\relax
\mciteBstWouldAddEndPuncttrue
\mciteSetBstMidEndSepPunct{\mcitedefaultmidpunct}
{\mcitedefaultendpunct}{\mcitedefaultseppunct}\relax
\EndOfBibitem
\bibitem{Bicudo:2015vta}
P.~Bicudo {\em et~al.}, \ifthenelse{\boolean{articletitles}}{\emph{{Evidence
  for the existence of $u d \bar{b} \bar{b}$ and the non-existence of $s s
  \bar{b} \bar{b}$ and $c c \bar{b} \bar{b}$ tetraquarks from lattice QCD}},
  }{}\href{https://doi.org/10.1103/PhysRevD.92.014507}{Phys.\ Rev.\
  \textbf{D92} (2015) 014507},
  \href{http://arxiv.org/abs/1505.00613}{{\normalfont\ttfamily
  arXiv:1505.00613}}\relax
\mciteBstWouldAddEndPuncttrue
\mciteSetBstMidEndSepPunct{\mcitedefaultmidpunct}
{\mcitedefaultendpunct}{\mcitedefaultseppunct}\relax
\EndOfBibitem
\bibitem{Francis:2016hui}
A.~Francis, R.~J. Hudspith, R.~Lewis, and K.~Maltman,
  \ifthenelse{\boolean{articletitles}}{\emph{{Lattice prediction for deeply
  bound doubly heavy tetraquarks}},
  }{}\href{https://doi.org/10.1103/PhysRevLett.118.142001}{Phys.\ Rev.\ Lett.\
  \textbf{118} (2017) 142001},
  \href{http://arxiv.org/abs/1607.05214}{{\normalfont\ttfamily
  arXiv:1607.05214}}\relax
\mciteBstWouldAddEndPuncttrue
\mciteSetBstMidEndSepPunct{\mcitedefaultmidpunct}
{\mcitedefaultendpunct}{\mcitedefaultseppunct}\relax
\EndOfBibitem
\bibitem{Karliner:2017qjm}
M.~Karliner and J.~L. Rosner,
  \ifthenelse{\boolean{articletitles}}{\emph{{Discovery of doubly-charmed
  $\Xi_{cc}$ baryon implies a stable ($b b \bar{u} \bar{d}$) tetraquark}},
  }{}\href{https://doi.org/10.1103/PhysRevLett.119.202001}{Phys.\ Rev.\ Lett.\
  \textbf{119} (2017) 202001},
  \href{http://arxiv.org/abs/1707.07666}{{\normalfont\ttfamily
  arXiv:1707.07666}}\relax
\mciteBstWouldAddEndPuncttrue
\mciteSetBstMidEndSepPunct{\mcitedefaultmidpunct}
{\mcitedefaultendpunct}{\mcitedefaultseppunct}\relax
\EndOfBibitem
\bibitem{Eichten:2017ffp}
E.~J. Eichten and C.~Quigg,
  \ifthenelse{\boolean{articletitles}}{\emph{{Heavy-quark symmetry implies
  stable heavy tetraquark mesons $Q_iQ_j \bar q_k \bar q_l$}},
  }{}\href{https://doi.org/10.1103/PhysRevLett.119.202002}{Phys.\ Rev.\ Lett.\
  \textbf{119} (2017) 202002},
  \href{http://arxiv.org/abs/1707.09575}{{\normalfont\ttfamily
  arXiv:1707.09575}}\relax
\mciteBstWouldAddEndPuncttrue
\mciteSetBstMidEndSepPunct{\mcitedefaultmidpunct}
{\mcitedefaultendpunct}{\mcitedefaultseppunct}\relax
\EndOfBibitem
\bibitem{Czarnecki:2017vco}
A.~Czarnecki, B.~Leng, and M.~B. Voloshin,
  \ifthenelse{\boolean{articletitles}}{\emph{{Stability of tetrons}},
  }{}\href{https://doi.org/10.1016/j.physletb.2018.01.034}{Phys.\ Lett.\
  \textbf{B778} (2018) 233},
  \href{http://arxiv.org/abs/1708.04594}{{\normalfont\ttfamily
  arXiv:1708.04594}}\relax
\mciteBstWouldAddEndPuncttrue
\mciteSetBstMidEndSepPunct{\mcitedefaultmidpunct}
{\mcitedefaultendpunct}{\mcitedefaultseppunct}\relax
\EndOfBibitem
\bibitem{Kiselev:2001fw}
V.~V. Kiselev and A.~K. Likhoded,
  \ifthenelse{\boolean{articletitles}}{\emph{{Baryons with two heavy quarks}},
  }{}\href{https://doi.org/10.1070/PU2002v045n05ABEH000958}{Phys.\ Usp.\
  \textbf{45} (2002) 455},
  \href{http://arxiv.org/abs/hep-ph/0103169}{{\normalfont\ttfamily
  arXiv:hep-ph/0103169}}, [Usp. Fiz. Nauk 172 (2002) 497]\relax
\mciteBstWouldAddEndPuncttrue
\mciteSetBstMidEndSepPunct{\mcitedefaultmidpunct}
{\mcitedefaultendpunct}{\mcitedefaultseppunct}\relax
\EndOfBibitem
\bibitem{Karliner:2014gca}
M.~Karliner and J.~L. Rosner,
  \ifthenelse{\boolean{articletitles}}{\emph{{Baryons with two heavy quarks:
  Masses, production, decays, and detection}},
  }{}\href{https://doi.org/10.1103/PhysRevD.90.094007}{Phys.\ Rev.\
  \textbf{D90} (2014) 094007},
  \href{http://arxiv.org/abs/1408.5877}{{\normalfont\ttfamily
  arXiv:1408.5877}}\relax
\mciteBstWouldAddEndPuncttrue
\mciteSetBstMidEndSepPunct{\mcitedefaultmidpunct}
{\mcitedefaultendpunct}{\mcitedefaultseppunct}\relax
\EndOfBibitem
\bibitem{Berezhnoy:2018bde}
A.~V. Berezhnoy, A.~K. Likhoded, and A.~V. Luchinsky,
  \ifthenelse{\boolean{articletitles}}{\emph{{Doubly heavy baryons at LHC}},
  }{}\href{http://arxiv.org/abs/1809.10058}{{\normalfont\ttfamily
  arXiv:1809.10058}}\relax
\mciteBstWouldAddEndPuncttrue
\mciteSetBstMidEndSepPunct{\mcitedefaultmidpunct}
{\mcitedefaultendpunct}{\mcitedefaultseppunct}\relax
\EndOfBibitem
\bibitem{Ali:2018xfq}
A.~Ali, Q.~Qin, and W.~Wang,
  \ifthenelse{\boolean{articletitles}}{\emph{{Discovery potential of stable and
  near-threshold doubly heavy tetraquarks at the LHC}},
  }{}\href{https://doi.org/10.1016/j.physletb.2018.09.018}{Phys.\ Lett.\
  \textbf{B785} (2018) 605},
  \href{http://arxiv.org/abs/1806.09288}{{\normalfont\ttfamily
  arXiv:1806.09288}}\relax
\mciteBstWouldAddEndPuncttrue
\mciteSetBstMidEndSepPunct{\mcitedefaultmidpunct}
{\mcitedefaultendpunct}{\mcitedefaultseppunct}\relax
\EndOfBibitem
\bibitem{Alves:2008zz}
LHCb collaboration, A.~A. Alves~Jr.\ {\em et~al.},
  \ifthenelse{\boolean{articletitles}}{\emph{{The \lhcb detector at the LHC}},
  }{}\href{https://doi.org/10.1088/1748-0221/3/08/S08005}{JINST \textbf{3}
  (2008) S08005}\relax
\mciteBstWouldAddEndPuncttrue
\mciteSetBstMidEndSepPunct{\mcitedefaultmidpunct}
{\mcitedefaultendpunct}{\mcitedefaultseppunct}\relax
\EndOfBibitem
\bibitem{LHCb-DP-2014-001}
R.~Aaij {\em et~al.}, \ifthenelse{\boolean{articletitles}}{\emph{{Performance
  of the LHCb Vertex Locator}},
  }{}\href{https://doi.org/10.1088/1748-0221/9/09/P09007}{JINST \textbf{9}
  (2014) P09007}, \href{http://arxiv.org/abs/1405.7808}{{\normalfont\ttfamily
  arXiv:1405.7808}}\relax
\mciteBstWouldAddEndPuncttrue
\mciteSetBstMidEndSepPunct{\mcitedefaultmidpunct}
{\mcitedefaultendpunct}{\mcitedefaultseppunct}\relax
\EndOfBibitem
\bibitem{Qiao:2012hp}
C.-F. Qiao, P.~Sun, D.~Yang, and R.-L. Zhu,
  \ifthenelse{\boolean{articletitles}}{\emph{{$B_c$ exclusive decays to
  charmonium and a light meson at next-to-leading order accuracy}},
  }{}\href{https://doi.org/10.1103/PhysRevD.89.034008}{Phys.\ Rev.\
  \textbf{D89} (2014) 034008},
  \href{http://arxiv.org/abs/1209.5859}{{\normalfont\ttfamily
  arXiv:1209.5859}}\relax
\mciteBstWouldAddEndPuncttrue
\mciteSetBstMidEndSepPunct{\mcitedefaultmidpunct}
{\mcitedefaultendpunct}{\mcitedefaultseppunct}\relax
\EndOfBibitem
\bibitem{LHCb-PAPER-2014-050}
LHCb collaboration, R.~Aaij {\em et~al.},
  \ifthenelse{\boolean{articletitles}}{\emph{{Measurement of $\Bcp$ production
  in proton-proton collisions at $\sqrt{s}=8$\tev}},
  }{}\href{https://doi.org/10.1103/PhysRevLett.114.132001}{Phys.\ Rev.\ Lett.\
  \textbf{114} (2015) 132001},
  \href{http://arxiv.org/abs/1411.2943}{{\normalfont\ttfamily
  arXiv:1411.2943}}\relax
\mciteBstWouldAddEndPuncttrue
\mciteSetBstMidEndSepPunct{\mcitedefaultmidpunct}
{\mcitedefaultendpunct}{\mcitedefaultseppunct}\relax
\EndOfBibitem
\bibitem{LHCb-PAPER-2017-042}
LHCb collaboration, R.~Aaij {\em et~al.},
  \ifthenelse{\boolean{articletitles}}{\emph{{Search for excited \Bc states}},
  }{}\href{https://doi.org/10.1007/JHEP01(2018)138}{JHEP \textbf{01} (2018)
  138}, \href{http://arxiv.org/abs/1712.04094}{{\normalfont\ttfamily
  arXiv:1712.04094}}\relax
\mciteBstWouldAddEndPuncttrue
\mciteSetBstMidEndSepPunct{\mcitedefaultmidpunct}
{\mcitedefaultendpunct}{\mcitedefaultseppunct}\relax
\EndOfBibitem
\bibitem{LHCB-PAPER-2011-044}
LHCb collaboration, R.~Aaij {\em et~al.},
  \ifthenelse{\boolean{articletitles}}{\emph{{First observation of the decay
  $\Bcp\to \jpsi\pip\pim\pip$}},
  }{}\href{https://doi.org/10.1103/PhysRevLett.108.251802}{Phys.\ Rev.\ Lett.\
  \textbf{108} (2012) 251802},
  \href{http://arxiv.org/abs/1204.0079}{{\normalfont\ttfamily
  arXiv:1204.0079}}\relax
\mciteBstWouldAddEndPuncttrue
\mciteSetBstMidEndSepPunct{\mcitedefaultmidpunct}
{\mcitedefaultendpunct}{\mcitedefaultseppunct}\relax
\EndOfBibitem
\bibitem{LHCb-PAPER-2013-063}
LHCb collaboration, R.~Aaij {\em et~al.},
  \ifthenelse{\boolean{articletitles}}{\emph{{Measurement of the $\Bcp$ meson
  lifetime using $\Bcp\to \jpsi\mup\neum X$ decays}},
  }{}\href{https://doi.org/10.1140/epjc/s10052-014-2839-x}{Eur.\ Phys.\ J.\
  \textbf{C74} (2014) 2839},
  \href{http://arxiv.org/abs/1401.6932}{{\normalfont\ttfamily
  arXiv:1401.6932}}\relax
\mciteBstWouldAddEndPuncttrue
\mciteSetBstMidEndSepPunct{\mcitedefaultmidpunct}
{\mcitedefaultendpunct}{\mcitedefaultseppunct}\relax
\EndOfBibitem
\bibitem{LHCb-PAPER-2014-060}
LHCb collaboration, R.~Aaij {\em et~al.},
  \ifthenelse{\boolean{articletitles}}{\emph{{Measurement of the lifetime of
  the $\Bcp$ meson using the $\Bcp\to \jpsi\pip$ decay mode}},
  }{}\href{https://doi.org/10.1016/j.physletb.2015.01.010}{Phys.\ Lett.\
  \textbf{B742} (2015) 29},
  \href{http://arxiv.org/abs/1411.6899}{{\normalfont\ttfamily
  arXiv:1411.6899}}\relax
\mciteBstWouldAddEndPuncttrue
\mciteSetBstMidEndSepPunct{\mcitedefaultmidpunct}
{\mcitedefaultendpunct}{\mcitedefaultseppunct}\relax
\EndOfBibitem
\bibitem{LHCb-PAPER-2010-002}
LHCb collaboration, R.~Aaij {\em et~al.},
  \ifthenelse{\boolean{articletitles}}{\emph{{Measurement of $\sigma(pp\to
  \bbbar X)$ at $\sqrt{s}=7$ TeV in the forward region}},
  }{}\href{https://doi.org/10.1016/j.physletb.2010.10.010}{Phys.\ Lett.\
  \textbf{B694} (2010) 209},
  \href{http://arxiv.org/abs/1009.2731}{{\normalfont\ttfamily
  arXiv:1009.2731}}\relax
\mciteBstWouldAddEndPuncttrue
\mciteSetBstMidEndSepPunct{\mcitedefaultmidpunct}
{\mcitedefaultendpunct}{\mcitedefaultseppunct}\relax
\EndOfBibitem
\bibitem{LHCb-PAPER-2011-018}
LHCb collaboration, R.~Aaij {\em et~al.},
  \ifthenelse{\boolean{articletitles}}{\emph{{Measurement of $\bquark$ hadron
  production fractions in 7\,TeV $\proton\proton$ collisions}},
  }{}\href{https://doi.org/10.1103/PhysRevD.85.032008}{Phys.\ Rev.\
  \textbf{D85} (2012) 032008},
  \href{http://arxiv.org/abs/1111.2357}{{\normalfont\ttfamily
  arXiv:1111.2357}}\relax
\mciteBstWouldAddEndPuncttrue
\mciteSetBstMidEndSepPunct{\mcitedefaultmidpunct}
{\mcitedefaultendpunct}{\mcitedefaultseppunct}\relax
\EndOfBibitem
\bibitem{LHCb-PAPER-2016-031}
LHCb collaboration, R.~Aaij {\em et~al.},
  \ifthenelse{\boolean{articletitles}}{\emph{{Measurement of the
  $\bquark$-quark production cross-section in 7 and 13\,TeV $\proton\proton$
  collisions}}, }{}\href{https://doi.org/10.1103/PhysRevLett.118.052002}{Phys.\
  Rev.\ Lett.\  \textbf{118} (2017) 052002}, Erratum
  \href{https://doi.org/10.1103/PhysRevLett.119.169901}{ibid.\   \textbf{119}
  (2017) 169901}, \href{http://arxiv.org/abs/1612.05140}{{\normalfont\ttfamily
  arXiv:1612.05140}}\relax
\mciteBstWouldAddEndPuncttrue
\mciteSetBstMidEndSepPunct{\mcitedefaultmidpunct}
{\mcitedefaultendpunct}{\mcitedefaultseppunct}\relax
\EndOfBibitem
\bibitem{LHCb-TDR-013}
LHCb collaboration, \ifthenelse{\boolean{articletitles}}{\emph{{LHCb VELO
  Upgrade Technical Design Report}}, }{}
  \href{http://cdsweb.cern.ch/search?p=CERN-LHCC-2013-021&f=reportnumber&action_search=Search&c=LHCb+Reports}
  {CERN-LHCC-2013-021}\relax
\mciteBstWouldAddEndPuncttrue
\mciteSetBstMidEndSepPunct{\mcitedefaultmidpunct}
{\mcitedefaultendpunct}{\mcitedefaultseppunct}\relax
\EndOfBibitem
\bibitem{LHCb-PAPER-2017-037}
LHCb collaboration, R.~Aaij {\em et~al.},
  \ifthenelse{\boolean{articletitles}}{\emph{{Measurement of the \Bpm
  production cross-section in $pp$ collisions at $\sqrt{s} = 7$ and $13$ TeV}},
  }{}\href{https://doi.org/10.1007/JHEP12(2017)026}{JHEP \textbf{12} (2017)
  026}, \href{http://arxiv.org/abs/1710.04921}{{\normalfont\ttfamily
  arXiv:1710.04921}}\relax
\mciteBstWouldAddEndPuncttrue
\mciteSetBstMidEndSepPunct{\mcitedefaultmidpunct}
{\mcitedefaultendpunct}{\mcitedefaultseppunct}\relax
\EndOfBibitem
\bibitem{Aaltonen:2007ah}
CDF collaboration, T.~Aaltonen {\em et~al.},
  \ifthenelse{\boolean{articletitles}}{\emph{{Observation of orbitally excited
  $B_s$ mesons}},
  }{}\href{https://doi.org/10.1103/PhysRevLett.100.082001}{Phys.\ Rev.\ Lett.\
  \textbf{100} (2008) 082001},
  \href{http://arxiv.org/abs/0710.4199}{{\normalfont\ttfamily
  arXiv:0710.4199}}\relax
\mciteBstWouldAddEndPuncttrue
\mciteSetBstMidEndSepPunct{\mcitedefaultmidpunct}
{\mcitedefaultendpunct}{\mcitedefaultseppunct}\relax
\EndOfBibitem
\bibitem{LHCb-PAPER-2012-030}
LHCb collaboration, R.~Aaij {\em et~al.},
  \ifthenelse{\boolean{articletitles}}{\emph{{First observation of the decay
  $B_{s2}^*(5840)^0 \to B^{*+}\Km$ and studies of excited $\Bs$ mesons}},
  }{}\href{https://doi.org/10.1103/PhysRevLett.110.151803}{Phys.\ Rev.\ Lett.\
  \textbf{110} (2013) 151803},
  \href{http://arxiv.org/abs/1211.5994}{{\normalfont\ttfamily
  arXiv:1211.5994}}\relax
\mciteBstWouldAddEndPuncttrue
\mciteSetBstMidEndSepPunct{\mcitedefaultmidpunct}
{\mcitedefaultendpunct}{\mcitedefaultseppunct}\relax
\EndOfBibitem
\bibitem{LHCb-PAPER-2013-044}
LHCb collaboration, R.~Aaij {\em et~al.},
  \ifthenelse{\boolean{articletitles}}{\emph{{Observation of the decay $\Bcp\to
  \Bs\pip$}}, }{}\href{https://doi.org/10.1103/PhysRevLett.111.181801}{Phys.\
  Rev.\ Lett.\  \textbf{111} (2013) 181801},
  \href{http://arxiv.org/abs/1308.4544}{{\normalfont\ttfamily
  arXiv:1308.4544}}\relax
\mciteBstWouldAddEndPuncttrue
\mciteSetBstMidEndSepPunct{\mcitedefaultmidpunct}
{\mcitedefaultendpunct}{\mcitedefaultseppunct}\relax
\EndOfBibitem
\bibitem{LHCb-PAPER-2015-047}
LHCb collaboration, R.~Aaij {\em et~al.},
  \ifthenelse{\boolean{articletitles}}{\emph{{Evidence for the
  strangeness-changing weak decay $\Xibm\to \Lb\pim$}},
  }{}\href{https://doi.org/10.1103/PhysRevLett.115.241801}{Phys.\ Rev.\ Lett.\
  \textbf{115} (2015) 241801},
  \href{http://arxiv.org/abs/1510.03829}{{\normalfont\ttfamily
  arXiv:1510.03829}}\relax
\mciteBstWouldAddEndPuncttrue
\mciteSetBstMidEndSepPunct{\mcitedefaultmidpunct}
{\mcitedefaultendpunct}{\mcitedefaultseppunct}\relax
\EndOfBibitem
\end{mcitethebibliography}

\end{document}